\documentclass{article}
\usepackage[twocolumn]{emulateapj}
\slugcomment{Accepted for publication in ApJ Letters}

\input psfig.sty

\newcommand\etal{{ et al. }}
\def\lsim{\mathrel{\rlap{\lower 4pt \hbox{\hskip 1pt $\sim$}}\raise 1pt \hbox
        {$<$}}}
\def\gsim{\mathrel{\rlap{\lower 4pt \hbox{\hskip 1pt $\sim$}}\raise 1pt \hbox
        {$>$}}}
 
\righthead{The Origin of Diversity of
Type Ia Supernovae and Environmental Effects}

\begin{document}
\title{The Origin of Diversity of
Type Ia Supernovae and Environmental Effects}

\author{Hideyuki Umeda, Ken'ichi Nomoto, Chiaki Kobayashi}
\affil{Department of Astronomy and Research Center for the Early Universe, 
University of Tokyo, Bunkyo-ku, Tokyo 113-0033, Japan \\ e-mail: 
umeda@astron.s.u-tokyo.ac.jp, nomoto@astron.s.u-tokyo.ac.jp,
chiaki@astron.s.u-tokyo.ac.jp}

\author{Izumi Hachisu}
\affil{Department of Earth Science and Astronomy, 
College of Arts and Sciences, University of Tokyo,
 Meguro-ku, Tokyo 153-8902, Japan \\ e-mail: 
hachisu@chianti.c.u-tokyo.ac.jp}

\author{Mariko Kato}
\affil{Department of Astronomy, Keio University, 
Hiyoshi, Kouhoku-ku, Yokohama 223-8521, Japan \\ e-mail: 
mariko@educ.cc.keio.ac.jp}

\begin{abstract}
 Observations suggest that the properties of 
Type Ia supernovae (SNe Ia) may depend on 
environmental characteristics, 
such as morphology, metallicity, and age of host galaxies.
The influence of these 
environmental properties on the resulting SNe Ia
is studied in this paper.
 First it is shown that the carbon mass fraction $X$(C) in 
the C+O white 
dwarf  SN Ia progenitors tends to be smaller for lower
metallicity and older the binary system age. It is then
suggested that the variation of $X$(C) causes the
diversity in the brightness of SNe Ia: a smaller $X$(C)
leads to a dimmer SN Ia. 
Further studies of the propagation of the 
turbulent flame are necessary 
to confirm this relation.  Our model for the SN Ia progenitors
then predicts that  when the progenitors  belong to an
older population or to a low metallicity environment, 
the number of bright SNe Ia is reduced, so that
the variation in brightness among the SNe Ia 
is also smaller. Thus our model can explain
why the mean SN Ia brightness and its dispersion depend
on the morphology of the host galaxies and on the distance 
of the SN from the center of the galaxy. It is further
predicted that at 
higher redshift ($z\gsim 1$) both the the mean brightness
of SNe Ia and its variation should be smaller in spiral 
galaxies than in elliptical galaxies. These variations 
are within the range observed 
in nearby SNe Ia. In so far as the variation in
$X$(C) is the most important cause for the
diversity among SNe Ia, the light curve shape method 
currently used to determine
the absolute magnitude of SNe Ia can be applied also
to high redshift SNe Ia.

\end{abstract}


\section{Introduction}

 It is widely accepted that Type Ia supernovae 
(SNe Ia) are thermonuclear
explosions of accreting C+O white dwarfs (WDs), although
the nature of the progenitor binary system and the detail of the
explosion mechanism are still under debate. 

 SNe Ia are good distance indicators, and provide a promising tool for 
determining cosmological parameters 
(e.g., Branch and Tammann 1992). 
From the observations of high redshift  
SNe Ia, both the SN Cosmology Project
(Perlmutter \etal 1999) and the High-z SN Search Team (Riess \etal 1998) 
have suggested a statistically significant
value for the cosmological constant. 
However, SNe Ia are not perfect standard candles, but show some
intrinsic variations in brightness. 
When determining the absolute peak luminosity 
of high-redshift SNe Ia, therefore,
these analyses have taken advantage of the empirical relation
existing between the peak brightness and the light curve shape (LCS). 
Since this relation has been obtained from
nearby SNe Ia only (Phillips 1993; Hamuy \etal 1995;
Riess, Press \& Kirshner 1995), 
it is important to examine whether it
depends systematically on environmental properties 
such as metallicity and age of the progenitor
system. This {\it Letter} addresses the issue of whether a difference
in the environmental properties is at the basis of the
observed range of peak brightness.

\bigskip

 There are some observational indications that SNe Ia are affected
by their environment. The most luminous SNe Ia seem to
occur only in spiral galaxies,
while both spiral and elliptical galaxies are hosts for dimmer
SNe Ia. Thus the mean peak brightness is dimmer in ellipticals than
in spiral galaxies (Hamuy \etal 1996). The SNe Ia rate per unit
luminosity at the present epoch 
is almost twice as high in spirals as in ellipticals 
(Cappellaro \etal 1997). Moreover,
Wang, H\"oflich \& Wheeler (1997) and Riess \etal (1999)
found that the variation of the 
peak brightness for SNe located in the outer regions in galaxies
is smaller.

H\"oflich, Wheeler, \& Thielemann (1998) examined how the
initial composition of the WD (metallicity and the C/O ratio)
affects the observed properties of SNe Ia. Umeda \etal (1999)
obtained the C/O ratio as a function of the main-sequence
mass and metallicity of the WD progenitors. In this  {\it Letter}
we suggest that the variation of the C/O ratio is the main
cause of the variation of SNe Ia brightness, with 
larger C/O ratio yielding brighter SNe Ia ($\S$ 2).
We then show that the C/O ratio 
depends indeed on environmental properties, such as the metallicity
 and age of the companion of the WD ($\S$ 3), and that 
our model can explain 
most of the observational trends discussed above ($\S 4$). We then make
some predictions about the brightness of SN Ia at higher 
redshift ($\S 5$).

\clearpage

\section{Explosion Model and the C/O Ratio of WD Progenitors}

 For the progenitors of SNe Ia, we adopt the single degenerate (SD)
Chandrasekhar mass model, in which an accreting C-O 
WD explodes when its mass reaches the
critical mass $M_{\rm Ia} \simeq 1.37-1.38 M_\odot$
(Nomoto, Thielemann, \& Yokoi 1984). 
Merging of white dwarfs is likely to
lead to accretion-induced-collapse rather than thermonuclear
explosion (Saio \& Nomoto 1998).
 Chandrasekhar mass models can reproduce well the spectrum and
the light curves of SNe Ia, assuming either the explosion is
induced by a deflagration or by a delayed detonation 
(H\"oflich \& Khokhlov 1996; Nugent \etal 1997).
In these models, the brightness of
SNe Ia is determined mainly by the mass of $^{56}$Ni synthesized
($M_{\rm Ni56}$). Observational data suggest that  $M_{\rm Ni56}$
for most SNe Ia  lies in the range $M_{\rm Ni56} \sim 0.4 - 0.8 M_\odot$ 
(e.g. Mazzali \etal 1998). This range of $M_{\rm Ni56}$
can result from differences in the C/O ratio in the progenitor WD
as follows.

 In the deflagration model, a faster propagation of the convective
deflagration wave results in a
larger $M_{\rm Ni56}$. For example, 
a variation of the propagation
speed by 15\% in the W6 -- W8 models results in 
$M_{\rm Ni56}$ values ranging between 0.5 and 
$0.7 M_\odot$ (Nomoto \etal 1984), which could explain the observations.
The actual propagation of the deflagration 
depends on the highly 
non-linear behavior of the turbulent flame (Niemeyer \& Hillebrand
1995), and so it may be very sensitive to the C/O ratio.
Qualitatively, a larger C/O ratio leads to the production 
of more nuclear energy and buoyancy force, thus leading to a faster 
propagation and a larger $M_{\rm Ni56}$.
Quantitatively, further studies of the turbulent flame are necessary
to confirm that the expected range of C/O results in the
required 15-20\% variation of the flame speed. 

In the delayed detonation model,  $M_{\rm Ni56}$ is 
predominantly determined by the deflagration-to-detonation-transition 
(DDT) density $\rho_{\rm DDT}$, at which the initially
subsonic deflagration turns into a 
 supersonic detonation (Khokhlov 1991).
 We reproduce the relation between $\rho_{\rm DDT}$ and $M_{\rm Ni56}$
in Figure 1 by 
performing hydrodynamical calculations
for several values of $\rho_{\rm DDT}$, as in 
Nomoto \etal (1997), Kishimoto \etal (1999), and Iwamoto \etal (1999). 
The pre-explosive WD model is 
model C6 (Nomoto \etal 1984), and the flame speed of the initial
deflagration is assumed to be 3\% of the local sound velocity.
Figure 1 shows that if the transition density varies in the range
 $\rho_{\rm DDT} \simeq 1.3 - 3 \times 10^7$ g cm$^{-3}$, 
the resulting variation
of  $M_{\rm Ni56}$ is large enough to explain the observations.


 Possible mechanisms for DDT to occur have been studied by
Arnett \& Livne (1994), Niemeyer \& Woosley (1997), and Khokhlov,
Oran \& Wheeler (1997): when the deflagration wave reaches a sufficiently
low density, $\rho \sim 10^7$ g cm$^{-3}$, 
 the turbulent motion
associated with the flame may destroy the burning front.
The resulting turbulent mixing between ashes and fuels efficiently 
heats up
the fuel and could produce a region with a very shallow temperature
gradient. In such a region, successive spontaneous ignitions
cause the over-driven deflagration to propagate
supersonically. This
may induce a detonation wave if the mass of the region exceeds
a critical mass $\Delta M_{\rm DDT}$.
This critical mass is quite sensitive to the carbon mass fraction
$X$(C), e.g. $\Delta M_{\rm DDT} \sim 10^{-19}$ and $10^{-14} M_\odot$
at $\rho = 3 \times 10^7$ g cm$^{-3}$ for $X$(C) = 1.0 and 0.5,
respectively (Niemeyer \& Woosley 1997).  Though the exact value of
$\rho_{\rm DDT}$ is still debated, and its dependence on $X$(C) has
not been studied, it is not unlikely that for a larger $X$(C) DDT can
occur at larger $\rho_{\rm DDT}$.  Hence a larger $X$(C) is likely to
result in a larger $\rho_{\rm DDT}$ and $M_{\rm Ni56}$.

In this {\it Letter}, therefore, 
we postulate that  $M_{\rm Ni56}$ and consequently brightness of
a SN Ia increase as the progenitors' C/O ratio increases,
as illustrated in Figure 1, where the range of
$M_{\rm Ni56} \sim 0.5-0.8 M_\odot$ is the 
result of an $X$(C) range $0.35-0.5$,
which is the range of $X$(C) values of our
progenitor models described below (Figure 2).
The $X$(C) -- $M_{\rm Ni56}$ relation we adopt is still only
a working hypothesis, which needs to be proved from studies
of the turbulent flame during explosion.

H\"oflich \etal (1998) considered the dependence on $X$(C)
but they assumed in their DDT model
that $\rho_{\rm DDT}$ and $X$(C) are independent parameters. 
They showed that
for the same $\rho_{\rm DDT}$ a smaller $X$(C) leads 
to a slightly brighter
SN Ia despite a slightly smaller $M_{\rm Ni56}$ produced,
because a smaller fraction of the explosion energy goes into the
kinetic energy. In this {\it Letter} 
we assume that $X$(C) is the 
primary parameter to determine $\rho_{\rm DDT}$ and thus $M_{\rm Ni56}$.
The assumed variation of $M_{\rm Ni56}$ is so large that 
a smaller $X$(C) yields an intrinsically dimmer SN.

\section{Metallicity and Age Effects}

 In this section we discuss
how the C/O ratio in the WD depends on the
metallicity and age of the binary system. 
 The C/O ratio in C+O WDs depends primarily on the main-sequence mass
of the WD progenitor and on metallicity. According to the evolutionary
calculations for 3$-$9 $M_\odot$ stars by Umeda \etal
(1999), the C/O ratio and its distribution 
are determined in the following evolutionary
stages of the close binary. 

 1) At the end of central He burning in the 3$-$9 
$M_\odot$ primary star, C/O$<1$ in
the convective core. The mass of the core is larger 
for more massive stars. 2) After central He exhaustion, the 
outer C+O layer
grows via He shell burning, where C/O$\gsim 1$ (Umeda \etal 1999). 
 3a) If the primary star becomes a red giant (case C evolution;
e.g. van den Heuvel 1994), it then undergoes the
second dredge-up, forming a thin He layer, and enters
the AGB phase. The C+O core mass, $M_{\rm CO}$,  at this phase
is larger for more massive stars. For a larger $M_{\rm CO}$
the total carbon mass fraction is smaller. 
4a) When it enters the AGB
phase, the star greatly expands and is assumed here to undergo Roche
lobe overflow (or a super-wind phase) and to form a C+O WD. Thus 
the initial mass of the WD, $M_{\rm WD}^{(0)}$, 
in the close binary at the beginning
of mass accretion is approximately equal to $M_{\rm CO}$.
3b) If the primary star becomes a He star (case BB evolution),
the second dredge-up in (3a) corresponds to the expansion of the
He envelope. 4b) The ensuing Roche lobe overflow again leads to a 
white dwarf of mass $M_{\rm WD}^{(0)}$ = $M_{\rm CO}$.

5) After the onset of mass accretion, 
the WD mass grows through steady H burning and weak He shell
flashes, as described in the WD wind model 
(Hachisu, Kato, \& Nomoto 1996, 1999, and Hachisu \etal 1999;
hereafter, HKN96, HKN99, and HKNU99, respectively).
The composition of the growing C+O layer is assumed to be C/O=1.
6) The WD grows in mass
and ignites carbon when its mass reaches $M_{\rm Ia} =1.367 M_\odot$, as 
in the model C6 of Nomoto \etal (1984).
Because of strong electron-degeneracy, carbon burning is unstable
and grows into a deflagration 
for a central temperature of $8\times
10^8$ K and a central
density of $1.47\times 10^9$ g cm$^{-3}$. At this stage, the
convective core extends to $M_r = 1.14M_\odot$ and the material is mixed 
almost uniformly, as in the C6 model.


In Figure 2 we show the carbon mass fraction $X$(C)
in the convective core of this pre-explosive WD, as a function
of metallicity ($Z$) and initial mass of the WD before the
onset of mass accretion, $M_{\rm CO}$.
 Figure 2 reveals that: 1) $X$(C) is smaller for larger $M_{\rm CO}$.
2) The dependence of $X$(C) on metallicity is small when plotted
against $M_{\rm CO}$, even though the relation between $M_{\rm CO}$ 
and the initial stellar mass depends sensitively on 
$Z$ (Umeda \etal 1999). 

{\it Metallicity dependent wind during mass accretion}:
     In the SD Chandrasekhar mass model for SNe Ia, a WD explodes as a
SN Ia only when its rate of the mass accretion ($\dot M$) is in a
certain narrow range (e.g., Nomoto \& Kondo 1991).  HKN96 showed that
the accreting WD blows a strong wind if $\dot M$ exceeds the rate
$\dot M_{\rm b}$ at which steady burning can process the accreted
hydrogen into He.  If the wind is sufficiently {\sl strong} (i.e., the
wind velocity $v_{\rm w}$ exceeds the escape velocity $v_{\rm esc}$ of
the WD), the WD can avoid the formation of a common envelope and
increase its mass continuously at a rate $\dot M_{\rm b}$ by blowing
the extra mass away in a wind.

     In this model, which is adopted in the present study, an
interesting metallicity effect has been found (Kobayashi \etal 1998;
Hachisu \& Kato 1999).  The wind velocity is higher for larger $M_{\rm
WD}$ and larger Fe/H because of higher luminosity and larger opacity,
respectively.  In order to blow sufficiently strong wind (i.e.,
$v_{\rm w} > v_{\rm esc}$), $M_{\rm WD}$ should exceed a certain mass
$M_{\rm w}$ (Fig. 6 of HKN99).  As seen in Figure 1 of Kobayashi \etal
(1998), $M_{\rm w}$ is larger for lower metallicity; e.g., $M_{\rm w}
=$ 0.65, 0.85, and 0.95 $M_\odot$ for $Z =$ 0.02, 0.01, and 0.004,
respectively. In order for a WD to grow its mass at $\dot M > \dot
M_{\rm b}$, its initial mass $M_{\rm CO}$ should exceed $M_{\rm w}$.
In other words, $M_{\rm w}$ is the metallicity-dependent minimum
$M_{\rm CO}$ required for a WD to become an SN Ia ({\sl strong wind
condition} in Fig.2).  The upper bound $M_{\rm CO} \simeq 1.07M_\odot$
is imposed by the condition that carbon should not ignite and is
almost independent of metallicity.

 As shown in Figure 2, the range of  $M_{\rm CO}$ can be converted 
into a range of $X$(C). From this
we find the following metallicity dependence for $X$(C):  
1) The upper bound of 
$X$(C), which is determined by the lower limit on $M_{\rm CO}$
imposed by the metallicity-dependent conditions for a strong wind, 
e.g.,  $X$(C) $\lsim 0.51$, 0.46 and 0.41, for $Z$=0.02, 0.01, and 0.004, 
respectively.  2) On the other hand,
the lower bound,  $X$(C) $\simeq 0.35-0.33$, 
does not depend much on $Z$, since it is imposed by the
maximum $M_{\rm CO}$.
 3) Assuming the relation between $M_{\rm Ni56}$ and $ X$(C) 
given in Figures 1 and 2, our model predicts the absence of brighter SNe Ia
in lower metallicity environment.

{\it Age effects}:
 In our model, the age of the progenitor system also constrains the
range of $X$(C) in SNe Ia. 
In the SD scenario, the lifetime of the binary system
is essentially the main-sequence lifetime of the companion star, 
which depends on its initial mass $M_2$. 
HKNU99 and HKN99 have obtained a constraint on $M_2$ by
calculating the evolution of accreting WDs
for a set of initial masses of the WD
($M_{\rm WD}^{(0)} \simeq M_{\rm CO}$) and of the companion ($M_2$),
and the initial binary period ($P_0$). In order for the WD mass
to reach $M_{\rm Ia}$, the donor star should transfer enough 
material at the appropriate accretion rates.
The donors of successful cases 
are divided into two categories: one is composed of
slightly evolved main-sequence stars with 
$M_2 \sim 1.7 - 3.6M_\odot$ (for $Z$=0.02),
and the other of red-giant stars with $M_2 \sim 0.8 - 3.1M_\odot$
(for $Z$=0.02) (HKN99, HKNU99; also Li \& van den Heuvel 1997). 

If the progenitor system is older than 2 Gyr, it should be a system
with a donor star of $M_2 < 1.7 M_\odot$ in the red-giant branch.  
Systems with $M_2 > 1.7 M_\odot$ become SNe Ia 
in a time shorter than 2 Gyr.
Likewise, for  a given age of the 
system,  $M_2$ must be
smaller than a limiting mass. This constraint on $M_2$ can be
translated into the presence of a minimum $M_{\rm CO}$ for a given
age, as follows:
For a smaller $M_2$, i.e. for the older system,
the total mass which can be transferred from the donor to the WD
is smaller. In order for $M_{\rm WD}$ to reach $M_{\rm Ia}$,
therefore, the initial mass of the WD,  $M_{\rm WD}^{(0)} \simeq
M_{\rm CO}$, should be larger.
This implies that the older system should have larger
minimum $M_{\rm CO}$ as indicated in Figure 2.
Using the $X$(C)$- M_{\rm CO}$  and
 $M_{\rm Ni56}- $X$(C)$ relations (Figs. 1 and 2), we conclude that
WDs in older progenitor systems have a smaller
$X$(C), and thus produce  dimmer SNe Ia.

\section{Comparison with Observations}

 The first observational indication which can be compared 
with our model is the possible 
dependence of the SN brightness on the morphology of the host galaxies.
Hamuy \etal (1996) found that 
the most luminous SNe Ia occur in spiral galaxies,
while both spiral and elliptical galaxies are hosts to dimmer
SNe Ia. Hence, the mean peak brightness is lower in elliptical than
in spiral galaxies. 

 In our model, this property is simply
understood as the effect of the different age of the companion. 
In spiral galaxies, star formation occurs continuously
up to the present time. Hence, both WD+MS and WD+RG systems can produce
SNe Ia. In  elliptical galaxies, on the other hand, star 
formation has long ended, typically more than 10 Gyr ago. 
Hence, WD+MS systems
can no longer produce SNe Ia. In Figure 3
 we show the  frequency of the expected SN I for a 
galaxy of mass $2 \times 10^{11} M_\odot$
for WD+MS and WD+RG systems
separately as a function of $M_{\rm CO}$. Here we 
use the results of HKN99 and HKNU99, and the
$M_{\rm CO}-X$(C) and
$M_{\rm Ni56}- X$(C) relations given in Figure 2.
Since a WD with smaller
$M_{\rm CO}$ is assumed to produce a brighter SN Ia (larger $M_{\rm Ni
56}$), our model predicts that dimmer SNe Ia occur both in 
 spirals and in ellipticals, while brighter ones occur only in spirals.
The mean brightness is smaller for
ellipticals and the total SN Ia rate per unit luminosity
is larger in spirals than in ellipticals.
These properties are consistent with observations.


 The second observational suggestion is the radial distribution of SNe Ia
in galaxies. Wang \etal (1997) and Riess \etal (1998) found that 
the variation of the 
peak brightness for SNe Ia located in the outer regions in galaxies 
is smaller. This behavior can be understood as the effect of metallicity.
As shown in Figure 2, even when
the progenitor age is the same, the minimum $M_{\rm CO}$ is larger for
a smaller metallicity because of the metallicity dependence 
of the WD winds. Therefore, our model predicts that the maximum brightness
of SNe Ia decreases as metallicity decreases.
Since the outer regions of galaxies are thought
 to have lower metallicities than the inner regions 
(Zaritsky, Kennicutt, Huchra 1994; Kobayashi \& Arimoto 1999),
our model is consistent with observations. Wang \etal
(1997) also claimed that SNe Ia may be deficient in the bulges of
spiral galaxies. This can be explained by the age effect, because the
bulge consists of old population stars. 

\section{Conclusions and Discussion}

 We have suggested that $X$(C) is the  quantity very likely to cause
the diversity in $M_{\rm Ni56}$ and thus in the brightness of SNe Ia.
 We have then shown that our model predicts that 
the brightness of SNe Ia depends on the environment, in a way
which is qualitatively consistent with the observations. 
Further studies of the propagation of the turbulent flame and the
DDT are necessary in order
to actually prove that $X$(C) is the key parameter.

 Our model predicts that when the progenitors belong to an
old population, or to a low metal environment, 
the number of very bright SNe Ia is small, so that
the variation in brightness is also smaller.
In spiral galaxies, the metallicity is significantly smaller
at redshifts $z\gsim 1$, and thus 
both the mean brightness of SNe Ia and its range tend to be smaller.
 At  $z\gsim 2$ SNe Ia would not occur in spirals at all
because the metallicity is too low.
In elliptical galaxies, on the other hand, the metallicity
at redshifts $z \sim 1-3$ is not very different from the present value.
However, the age of the galaxies at $z\simeq 1$ is only about 5 Gyr,
so that the mean brightness of SNe Ia and its range
tend to be larger at $z\gsim 1$ than in the present
ellipticals because of the age effect. 

 We note that 
the variation of $X$(C) is larger in metal-rich nearby
spirals than in high redshift galaxies.  Therefore, if $X$(C) is the
main parameter responsible for the diversity of SNe Ia, and if the LCS
method is confirmed by the nearby SNe Ia data, the LCS method can also
be used to determine the absolute magnitude of high redshift SNe Ia.

 We wish to thank P. Mazzali, K. Iwamoto and N. Kishimoto
for useful discussion and suggestions.
This work has been supported in part by the grant-in-Aid for
Scientific Research (0980203, 09640325) and COE research
(07CE2002) of the Ministry of Education, Science, Culture 
and Sports in Japan.


\begin{figure*}[h]
\psfig{figure=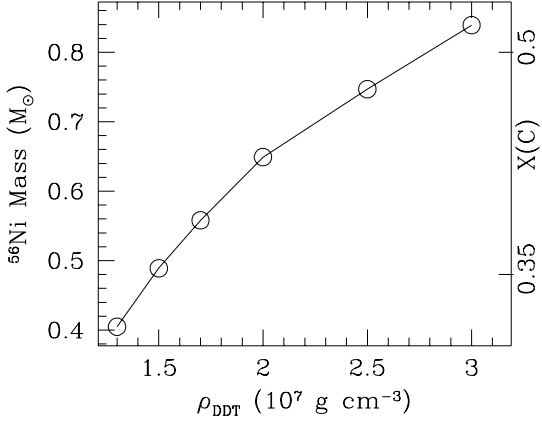,width=10.0cm}
\caption{
Deflagration to detonation transition density ($\rho_{\rm
DDT}$) vs. $^{56}$Ni mass ($M_{\rm Ni56}$).
The $X$(C) axis is approximately given as discussed in $\S$ 2.
These results are obtained for $X$(C)=0.43. For a given $\rho_{\rm DDT}$
varying $X$(C)
between $X$(C)=0.35 $-$ 0.50 changes the $^{56}$Ni mass only by
$\sim \pm 0.02 M_\odot$.
\label{FIG1}}
\end{figure*}

\begin{figure*}[h]
\psfig{figure=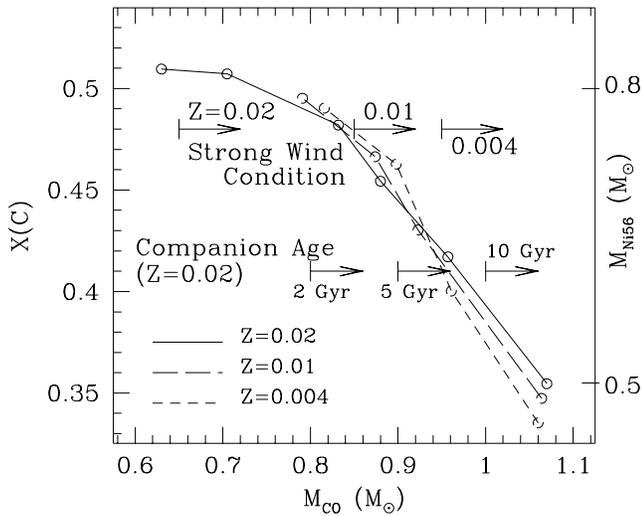,width=10.0cm}
\caption
{ The total  $^{12}$C mass fraction
 included in the convective core of mass, $M=1.14M_\odot$, just
before the SN Ia explosion
 as a function of the C+O core mass before the onset of mass
accretion, $M_{\rm CO}$. The lower bounds of $M_{\rm CO}$  
obtained from the age effects and the conditions 
for strong wind to blow are
also shown by arrows. The axis of $M_{\rm Ni56}$ is obtained 
from the $X$(C) $- M_{\rm Ni56}$ relation assumed in Figure 1.
\label{FIG2}}
\end{figure*}

\begin{figure*}[h]
\psfig{figure=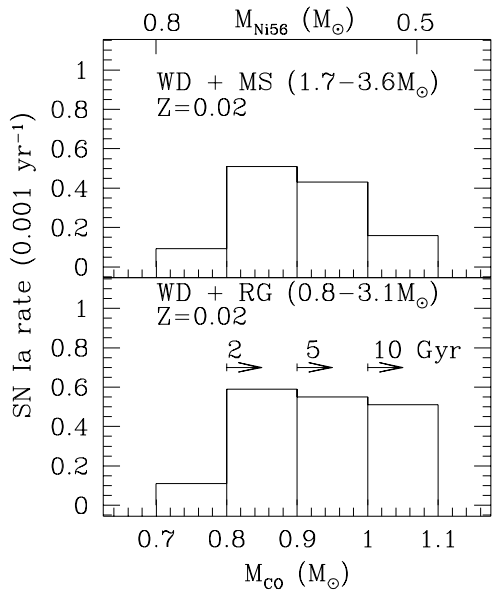,width=13.0cm}
\caption
{ SN Ia frequency for a galaxy of mass $2 \times 10^{11} M_\odot$
as a function of $M_{\rm CO}$ for $Z$=0.02.
For the WD+RG system, constraints from the companion's age are
shown by the arrows. SNe Ia from the WD+MS system occur in spirals
but not in ellipticals because of the age effect. $M_{\rm CO}$ and
$M_{\rm Ni56}$ can be related as shown here if the $X$(C) $- M_{\rm Ni56}$
relation in Figure 1 is adopted.
\label{FIG3}}
\end{figure*}

\end{document}